\documentstyle[12pt]{article}

\begin{document}

\begin{center}
{\Large{\bf The Abelian Topological Mass Mechanism \\
>From Dimensional Reduction}} \\ [14mm]
{\bf Adel Khoudeir}\footnote { e-mail: adel@ciens.ula.ve}\\ 
{\it Centro de Astrof\'{\i}sica Te\'orica, Departamento de F\'{\i}sica,
Facultad de Ciencias, Universidad de los Andes, M\'erida, 5101,
Venezuela.}\\[4mm] 

{\bf Abstract}
\end{center}

We show that the abelian topological mass mechanism 
in four dimensions, described by the Cremmer-Sherk action, 
can be obtained from dimensional reduction in five dimensions. 
Starting from a gauge invariant action in five dimensions, 
where the dual equivalence between a massless 
vector field and a massless second-rank antisymmetric field 
in five dimensions is established, the dimensional reduction 
is performed keeping only one massive mode. Furthermore, the 
Kalb-Ramond action and the Stuckelberger formulation for 
massive spin-1 are recovered.

\newpage

Several alternatives of the Higgs Mechanism, 
based on the coupling of vector fields with antisymmetric fields through 
topological terms, have been developed in the last 
years. In particular, the abelian Cremmer-Sherk 
theory \cite{cs} has been studied extensively \cite{ext} as 
the prototype of this proposal. Its non-abelian extension is a 
Freedman-Townsend theory\cite{ft} that can be derived using the 
self-interaction mechanism \cite{adel}. Also, other attempts 
to search a non-abelian generalization have been formulated
\cite{varios}, however, serious problems with renormalizability in all 
these non-abelian generalizations has been pointed out in 
reference \cite{henneaux}. 
An interesting aspect is the fact that the Cremmer-Sherk theory 
is related by duality\cite{siva}, \cite{adel} with the Kalb-Ramond
theory 
\cite{kr}, where this latter can be obtained by dimensional reduction of 
the second-rank antisymmetric theory in five dimensions 
keeping only one massive mode \cite{govi}, \cite{bn}. 
It is worth recalling that the Kalb-Ramond theory 
provides mass in a non-topological way and can be understood as the 
resulting theory after condensation of magnetic monopoles in 
four dimensions QED\cite{quetru}
In this letter, we will show that it is possible to obtain the 
Cremmer-Sherk theory in four dimensions by dimensional reduction, 
despite Barcelos-Neto has claimed 
that the Cremmer-Sherk theory can not come from dimensional reduction of 
any gauge theory in five dimensions\cite{bn}. 

Let us describe briefly the dual equivalence between the Cremmer-Sherk 
and Kalb-Ramond theories in four dimensions. The Cremmer-Sherk action, 
which provides mass to spin-1 fields without spoil gauge invariance, 
is written down as 
\begin{equation}
I_{CrSc} = \int d^4 x [-\frac{1}{4}F_{mn}F^{mn} -
\frac{1}{12}H_{mnp}H^{mnp} 
- \frac{1}{4}\mu\epsilon^{mnpq}B_{mn}F_{pq}],
\end{equation}
where $F_{mn} \equiv \partial_m A_n - \partial_n A_m$ and 
$H_{mnp} \equiv \partial_m B_{np} + \partial_n B_{pm} + \partial_p
B_{mn}$ 
are the field strengths associated with the $A_m$ and $B_{mn}$ fields. 
This action is invariant under gauge transformations: $\delta A_m = 
\partial_m \lambda$ and $\delta B_{mn} = \partial_m \zeta_n - 
\partial_n \zeta_m$. We observe that this action has global symmetries, 
for instance: $A_m \rightarrow A_m - \epsilon_m$, with $\epsilon_m$ the
global parameter. An useful way to obtain the dual theory relies on 
gauging this local symmetry \cite{quev} by introducing an antisymmetric 
field $a_{mn}$, such that the field strenght $F_{mn}$ is modified by 
$F_{mn} \equiv \partial_m A_n - \partial_n A_m - \frac{1}{2}(a_{mn} - 
a_{nm})$ and adding a BF term, which assures the non-propagation of 
$a_{mn}$. Then, we have the following action
\begin{eqnarray}
I_M &=& \int d^4 x [-\frac{1}{4}[\partial_m A_n - \partial_n A_m + 
\frac{1}{2}(a_{mn} - a_{nm})]^2 - \frac{1}{12}H_{mnp}H^{mnp}\\
&-& \frac{1}{4}\mu\epsilon^{mnpq}B_{mn}[\partial_m A_n - \partial_n A_m
+ 
\frac{1}{2}(a_{mn} - a_{nm})] + \frac{1}{2}a_m \epsilon^{mnpq}\partial_n 
a_{pq}]\nonumber
\end{eqnarray}
and we now have  new gauge symmmetries, given by: $\delta A_m = 
-\epsilon_{m(x)}$, $\delta a_m = \partial_m\alpha$ and $\delta a_{mn} = 
\partial_{m}\epsilon_{n} - \partial_{n}\epsilon_{m}$, which allow us 
to fix the gauge $A_{m} = 0$. After fixing this gauge, the action
becomes
\begin{equation}
I = \int d^4 x [-\frac{1}{4}a_{mn}a^{mn} - \frac{1}{12}H_{mnp}H^{mnp} - 
\frac{1}{4}\mu\epsilon^{mnpq}B_{mn}a_{pq} + \frac{1}{2}\epsilon^{mnpq}
a_{mn}\partial_p a_q].
\end{equation}
Integrating out the field $a^{mn}$($=-\frac{1}{2}\epsilon^{mnpq}
[\mu B_{pq} - (\partial_p a_q -  \partial_q a_p)]$, the Kalb-Ramond 
action is obtained:
\begin{equation}
I_{KR} = \int d^4 x [-\frac{1}{12}H_{mnp}H^{mnp} - \frac{1}{4}
[(\partial_m a_n -  \partial_n a_m) - \mu B_{mn}]
[(\partial^m a^n -  \partial^n a^m) - \mu B_{mn}]].
\end{equation}
This action is invariant under gauge transformations given by: 
$\delta B_{mn} = \partial_m \zeta_n - \partial_n \zeta_m$ and 
$\delta a_m = \partial_m\alpha - \mu\zeta_m$, which allow us 
gauged away the $a_m$ field, leading to the massive $B_mn$ 
antisymmetric action, describing three masive degrees of freedom 
like abelian massive vector theories in four dimensions.
Having seen that in four dimensions the Cremmer-Sherk and Kalb-Ramond 
theories are equivalent by duality and that the Kalb-Ramond can be
obtained 
from dimensional reduction of a 2-form in five dimensions, it must be
possible 
to obtain the Cremmer-Sherk from dimensional reduction of some gauge
theory 
in five dimensions. 

Since the dual of a 2-form in five dimensions is a 
1-form, we will consider the following master action in five dimensions
(the indices $M,N$ run over $0,1,...4$)
\begin{equation}
I = \int d^5 x [\frac{1}{4}V^{MN}V_{MN} -
\frac{1}{2}V^{MN}[\partial_{M}A_{N} - 
\partial_{N}A_{M}]],
\end{equation}
which is just the first order formulation of the Maxwell action and 
where $V^{MN}$ and $A_{M}$ are considered as independent variables. 
Indeed, the Maxwell action is recovered
after eliminating $V^{MN}$ through its equation of motion($V_{MN} = 
[\partial_{M}A_{N} - \partial_{N}A_{M}]$). 
On the other hand, the equations of motion after independent variations
in 
$A_{M}$ are
\begin{equation}
\partial_M V^{MN} = 0,
\end{equation}
whose solution in five dimensions can be written(locally) as
\begin{equation}
V^{MN} = \frac{1}{2}\epsilon^{MNPQR}\partial_{P}B_{QR} \equiv 
\frac{1}{6}\epsilon^{MNPQR}H_{PQR}.
\end{equation}
Substituting this expression for $V^{MN}$ into eq. (5), we obtain the 
action for the second-rank antisymmetric field $I_B = - \frac{1}{12}
H_{MNP}H^{MNP}$. 
In general, for $D > 2$ dimensions, the solution of eq. (6) is
\begin{equation}
V^{MN} = \frac{1}{(D-3)!}\epsilon^{MNPQ_1 ...Q_{D-3}} \partial_{P} 
B_{Q_1 ...Q_{D-3}}
\end{equation}
and going back into eq. (5), the 
usual duality between 1 and 0-forms in three dimensions, 1 and 1-forms 
in four dimensions and so on, is achieved.

Now, we are going to perform the dimensional reduction. 
Let us deal with complex fields in five dimensions, then we consider the 
following action
\begin{equation}
I = \int d^5 x [\frac{1}{4}V^{MN}V_{MN}^{*} - 
\frac{1}{4}V^{MN*}[\partial_{M}A_{N} - \partial_{N}A_{M}] - 
\frac{1}{4}V^{MN}[\partial_{M}A_{N}^{*} - \partial_{N}A_{M}^{*}]].
\end{equation}
Using the method of dimensional reduction developed in \cite{csscha}
for generating mass, we write in the limit $R \rightarrow 0$
($R$ being the radius of the small circle in the coordinate $x^4$ where
we are 
compactifying) 
\begin{eqnarray}
V_{mn(x,x^{4})} &=& F_{mn(x)}e^{i\mu x^4}, \quad A_{m(x,x^{4})} = 
A_{m(x)}e^{i\mu x^4},\\
V_{m4(x,x^{4})} &=& i V_{m(x)}e^{i\mu x^4}, \quad A_{4(x,x^{4})} = 
i\phi_{(x)} e^{i\mu x^4}.\nonumber
\end{eqnarray}
In consequence, we obtain the following reduced action to four
dimensions.
\begin{equation}
I_{red4D} = \int d^4 x [\frac{1}{4}F^{mn}F_{mn} - \frac{1}{2}F^{mn}
[\partial_m A_n - \partial_n A_{m}] + \frac{1}{2}V^m V_m - 
V^{m}[\partial_m\phi - \mu A_m]].
\end{equation}
>From this reduced action, we can eliminate the $F_{mn}$ and $\phi$
fields 
through its equations of motion: 
\begin{equation}
F_{mn} = \partial_m A_n - \partial_n A_{m}, \quad \partial_m V^m = 0
\end{equation}
and we have the following solution for $V^m$ in four dimensions
\begin{equation}
V^m = -\frac{1}{2}\epsilon^{mnpq}\partial_{n}B_{pq} = -\frac{1}{6}
\epsilon^{mnpq}H_{npq}.
\end{equation}
Putting this back into the reduced action, the Cremer-Sherk action is
obtained.
 
Furthermore, the field equation after varying the $A_m$ field is 
\begin{equation}
\partial_n F^{nm} = -\mu V^m,
\end{equation}
which combined with eq. (13), lead us to following solution 
for $F^{mn}$
\begin{equation}
F^{mn} = \frac{1}{2}\epsilon^{mnpq}[-\mu B_{pq} + (\partial_p a_q - 
\partial_q a_p]
\end{equation}
Then, we can eliminate the $\phi$ and $A_m$ fields 
and substituting eqs. (13) and (15) into the reduced action,  
the Kalb-Ramond action is obtained. 
Moreover, from the reduced action we can integrate out the $V_m$ field 
($V_m = \partial_m\phi - \mu A_m$), in order to reach the Stuckelberger 
formulation for massive spin-1 fields: $I_{Stuck} = \int d^4 x 
[-\frac{1}{4}F^{mn}F^{mn} - \frac{1}{2}\mu^2 
(A_m - \frac{1}{\mu}\partial_m\phi)^2]$.
 
Summarizing, we have shown that the Cremmer-Sherk action, which 
provides mass for vector fields compatible with gauge symmetry, 
as well as the Kalb-Ramond and the Stuckelberger actions, 
can be obtained from dimensional reduction of a gauge theory, 
which reflects the dual equivalence between 
the Maxwell field and the second-rank antisymmetric field in five 
dimensions.

\begin{center}
{ACKNOWLEDGEMENTS}
\end{center}

The author would like to thank to the Consejo de Desarrollo 
Cient\'{\i}fico y Human\'{\i}stico de la Universidad de los
Andes(CDCHT-ULA) 
by institutional support under project C-862-97.

\end{document}